\documentclass[prb,twocolumn,superscriptaddress]{revtex4}

\usepackage{graphicx}
\usepackage{bm}

\def\be{\begin{equation}}
\def\ee{\end{equation}}
\def\ba{\begin{eqnarray}}
\def\ea{\end{eqnarray}}

\begin{document}

\title{
Nernst effect, quasiparticles, and $d$-density waves
in cuprates}

\author{V. Oganesyan}
\affiliation{Department of Physics, Princeton University,
Princeton, New Jersey 08544, USA}
\author{Iddo Ussishkin}
\affiliation{Theoretical Physics Institute, University of Minnesota,
Minneapolis, Minnesota 55455, USA}

\date[]{22 December 2003}

\begin{abstract}
We examine the possibility that the large Nernst signal observed in the
pseudogap regime of hole-doped cuprates originates from quasiparticle
transport in a state with $d$-density wave (DDW) order, proposed by S.
Chakravarty \emph{et al.}\ [Phys.\ Rev.\ B \textbf{63}, 094503 (2001)].
We find that the Nernst coefficient can be moderately enhanced in
magnitude by DDW order, and is generally of negative sign. 
Thus, the quasiparticles of the DDW state
cannot account for the large and positive
Nernst signal observed in the pseudogap phase of the cuprates. However,
the general considerations outlined in this paper may be of broader
relevance, in particular to the recent measurements of Bel \emph{et
al.}\ in NbSe$_2$ and CeCoIn$_5$ [Phys.\ Rev.\ Lett.\ \textbf{91}, 066602 (2003);
\emph{ibid.}\ \textbf{92}, 217002 (2004)].
\end{abstract}

\maketitle

Much of the original interest in the pseudogap phenomena in
high-temperature superconductors (HTSC) stemmed from the belief that it
represented, in some way, a vestige of the superconducting
state,~\cite{rvbpgapetc} and thus could offer insights as to the
latter's origins. In recent years, however, fluctuations of other
orders, such as spin and charge density waves, have been detected in
some of these materials,~\cite{rvbpgapetc,fluctorder} suggesting that a proper
understanding of the pseudogap phase requires incorporating these
ordering instabilities. A proposal by Chakravarty \emph{et
al.}~\cite{DDW-proposal} goes one step further: according to them the
pseudogap is a consequence of long range $d$-density wave (DDW) order,
a pattern of circulating currents. At the mean-field level, this
unconventional density wave order possesses quasiparticles, and much of
its potential for explaining pseudogap phenomena stems from the changes
in the single particle spectrum due to the breaking of translational
symmetry.~\cite{sharp-signature,DDW-ARPES,DDW-infrared,boebinger,Rpez,mook,rigal}

The discovery~\cite{Ong-nature} of a large Nernst effect in hole-doped
cuprates posed a new major challenge for the theoretical description of
the pseudogap regime. The Nernst effect (see Sec.~\ref{sec:1theory}) is
anomalously large in the non-superconducting state of underdoped
samples, extending to rather high temperatures above
$T_c$.~\cite{Ong-nature,Ong-onset,Capan,Wen} This surprising result
contrasts with conventional materials where the effect is generally
small in the normal state.

In superconductors (both conventional and high-temperature), a large
Nernst signal is observed below $T_c$ as field induced vortices become
depinned and float down the thermal gradient, their motion producing a
transverse voltage by phase slips.~\cite{Huebener} By continuity,
therefore, the observed signal in the pseudogap may be associated with
collective fluctuations of the superconducting order parameter. In this
spirit, Ong and collaborators have interpreted their results as
evidence for vortices above $T_c$.~\cite{Ong-nature} Recently, a
detailed analysis by Sondhi, Huse, and one of us~\cite{Iddo} has shown
that superconducting fluctuations can produce a sizable effect in the
cuprates (see also Ref.~\onlinecite{Subroto}). Other
works~\cite{Kontani,Chicago,Patrick} have also suggested, in one way or
another, that the Nernst effect is a result of collective phenomena of
superconducting origin.

In contrast, here we consider whether single particle transport can be
a source of an enhanced Nernst signal. This is done first on rather
general terms, which should be applicable to different systems
(including spin and charge density waves). We then consider in detail
whether the onset of DDW order can by itself account for the
experimental observations in HTSC. We also discuss two experiments by
Bel and collaborators, in which a large Nernst signal is observed in
the normal state of NbSe$_2$,~\cite{ambipolar} and, very recently, in a
heavy fermion compound CeCoIn$_5$.~\cite{kondo}

We begin by re-examining the conventional theory of transport in metals
in Sec.~\ref{sec:1theory}, focusing in particular on the Nernst
phenomena it predicts.
In Sec.~\ref{sec:ddw} we compute the effects of the DDW order. Experimental results are
discussed in the following Section. We
close with a short Summary and an Appendix.

\section{General considerations}
\label{sec:1theory}

The Nernst effect is the off-diagonal component of the thermopower
tensor $Q$, measured in the absence of electric flow and with magnetic
field $B$ in the $\mathbf{\hat z}$ direction (in the cuprates,
perpendicular to the copper-oxide planes). The thermopower tensor is
given by
\begin{equation}
\mathbf{E} = Q \cdot \bm{\nabla} T = \sigma^{-1} \cdot \alpha \cdot
\bm{\nabla} T,
\end{equation}
where $\sigma$ and $\alpha$ are the conductivity and Peltier
(thermoelectric) conductivity tensors, respectively. The relation
$Q_{xy} = -Q_{yx}$ generally holds only when the system has isotropic
transport tensors (e.g., $\sigma_{xx} = \sigma_{yy}$). In this case,
the Nernst signal is given by
\begin{equation} \label{eq:1Q_xy}
Q_{xy} = \frac{E_y}{(- \nabla T)_x} = \frac{\alpha_{xy} \sigma_{xx}
-\alpha_{xx} \sigma_{xy}}{\sigma_{xx}^2+\sigma_{xy}^2} .
\end{equation}
Here, we follow the sign convention such that the Nernst signal arising
from vortex flow in a superconductor is positive. We will be mainly
concerned with situations where $Q_{xy} \propto B$ over a wide range of
fields, and therefore concentrate on the calculation of the Nernst
coefficient $\nu=Q_{xy}/B$.

Quasiparticle contribution to the Nernst effect is usually argued to be
small, as it is strictly zero in the simple Drude model due to the
``Sondheimer cancellation''~\cite{Ong-onset} between the two terms in
Eq.~(\ref{eq:1Q_xy}). Generally, in any realistic system, such a
cancellation will be incomplete. In this Section we delineate the
factors determining the magnitude of $Q_{xy}$ under general conditions
of validity of the Boltzmann theory of transport.~\cite{z2,allen}

Solving the Boltzmann equation at low temperature $T$, the thermoelectric
tensor $\alpha$ is related to the conductivity tensor $\sigma$ through
\begin{equation} \label{eq:1alpha}
\alpha = - \frac{\pi^2}{3} \, \frac{k_B^2 T}{e} \, \frac{\partial
\sigma}{\partial \mu} ,
\end{equation}
where $\mu$ is the chemical potential, and $-e<0$ is the electron
charge.~\cite{dmu} Using Eq.~(\ref{eq:1Q_xy}), the Nernst coefficient
of a degenerate Fermi liquid may then be reduced to
\begin{equation} \label{eq:1nernst1}
\nu = -\frac{\pi^2}{3} \, \frac{k_B^2 T}{e B} \, \frac{\partial
\Theta_H}{\partial \mu}.
\end{equation}
Here, $\Theta_H = \sigma_{xy} / \sigma_{xx}$ is the Hall angle
 to linear order in magnetic field.
The Nernst coefficient is thus encoded
in the dependence of $\Theta_H$ on $\mu$.

We proceed by assuming a constant scattering time $\tau$, in order to
focus on the role of Fermi surface geometry in the Nernst effect,
returning to the details of $\tau$ below. In two dimensions and to
linear order in magnetic field the longitudinal and Hall conductivities
are expressed in terms of integrals over the Fermi surface,
\begin{equation} \label{eq:1sigma-xx}
\sigma_{xx} = e^2 \tau \int \frac{d^2 k}{(2 \pi)^2} \, v_x^2 \, \delta
(\epsilon_\mathbf{k} - \mu) ,
\end{equation}
\begin{equation} \label{eq:1sigma-xy}
\sigma_{xy} = - \frac{e^3 B \tau^2}{\hbar c} \int \frac{d^2 k}{(2
\pi)^2} \left( v_x^2 \frac{\partial v_y}{\partial k_y} - v_x v_y
\frac{\partial v_y}{\partial k_x} \right) \delta (\epsilon_\mathbf{k} -
\mu) ,
\end{equation}
where $\mathbf{v} = (\partial \epsilon_\mathbf{k} / \partial
\mathbf{k}) / \hbar$ is the velocity of the quasiparticle (and spin and
band indices are suppressed).~\cite{NPOgeom}

It is convenient to rearrange the expression for the Nernst
coefficient, Eq.~(\ref{eq:1nernst1}), as
\begin{equation} \label{eq:1simplenu}
B \nu = - \frac{\pi^2}{3} \, \frac{k_B}{e}  \, \frac{a^2}{\ell_B^2} \,
\frac{k_B T \tau}{\hbar} \,\frac{\partial \Upsilon }{\partial \mu} .
\end{equation}
Here, $k_B / e \approx 86 \, \mu \text{V} / \text{K}$ is the only
dimensionful factor, setting the natural scale for a thermopower
measurement. The reduction factor involving the lattice spacing [$a$]
and the magnetic length [$\ell_B = (\hbar c / e B)^{1/2}$] encodes the
field's weakness on a scale natural to the system.  This is multiplied
(apart from the numerical factor) by the ratio of the thermal energy
and relaxation rate, and by the dimensionless derivative $\partial
\Upsilon / \partial \mu$. The energy scale $\Upsilon$ is constructed by
formally stripping the weak-field Hall angle of its dependence on
magnetic field and relaxation,
\begin{equation} \label{eq:1Upsilon}
\Upsilon=\frac{\ell_B^2}{a^2} \, \frac{\hbar}{\tau} \Theta_H .
\end{equation}
For an isotropic Fermi surface of holes $\Theta_H = \omega_c \tau$
(with the cyclotron mass given by $m = \hbar k_F / v_F$), and $\Upsilon
= \hbar^2 / m a^2$. Finally, it is important to note that when
$\partial \Upsilon / \partial \mu > 0$ the Nernst coefficient is
negative.

\begin{figure}
\includegraphics[width=3.25in]{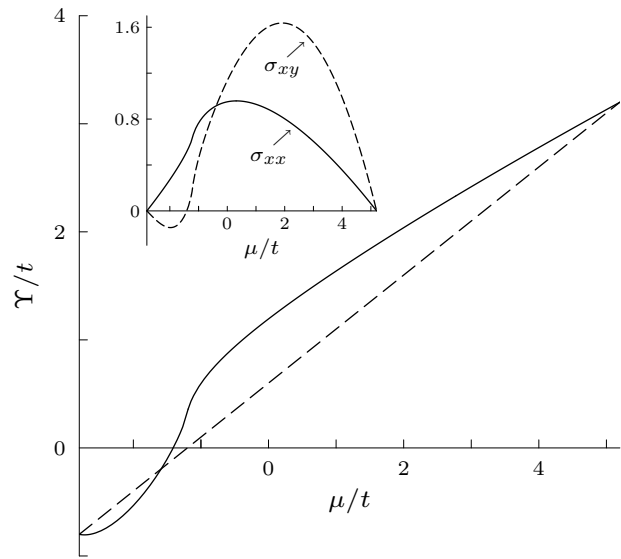}
\caption{\label{fig:whole-band} The energy scale $\Upsilon$ (Hall angle
in appropriate units) is compared against the naive estimate of a
uniform slope over the entire band range, from 0 to 2 electrons per
site (using the same hopping parameters as in Sec.~\ref{sec:ddw}, $t'
= 0.3 t$; half filling is at $\mu = -0.66 t$). In the inset we
present $\sigma_{xx}$ and $\sigma_{xy}$ in units $(e^2 / \hbar) (t \tau
/ \hbar)$ and $(e^2 / \hbar) (t \tau / \hbar)^2 (a^2 / \ell_B^2)$
respectively.}
\end{figure}

We now consider a tight-binding model on the square lattice with
nearest and next-nearest neighbor hopping parameters ($t$-$t'$ model)
as a simple, concrete example with which to explore
Eq.~(\ref{eq:1simplenu}), and also for future use in Sec.~\ref{sec:ddw}.
The free electron expression for $\Upsilon$ suggests a simple way of
approximating the slope of $\Upsilon (\mu)$ by its average over the
entire band,
\begin{equation} \label{eq:1mmW}
\overline{\frac{\partial \Upsilon}{\partial \mu}} = \frac{1}{W} \,
\frac{\hbar^2}{a^2} \left( \frac{1}{m_e} + \frac{1}{m_h} \right) =
\frac{1}{2}.
\end{equation}
Here, $m_e$ ($m_h$) is the electron (hole) mass near the bottom (top)
of the band, $W$ is the overall bandwidth, and the result is
independent of the hopping parameters. Remarkably, this crude estimate
is exceedingly accurate (see Fig.~\ref{fig:whole-band}) and for
practical purposes $\partial \Upsilon/\partial \mu$ is a number of
order 1.~\cite{vanHove} With $\partial \Upsilon / \partial \mu \approx
1$, the estimate for the Nernst coefficient then boils down to
\begin{equation} \label{eq:1nu-estimate}
\nu \approx - 100 \, \frac{k_B T \tau}{\hbar} \,
\frac{\text{nV}}{\text{K} \, \text{T}},
\end{equation}
where the last expression is evaluated using
$a=5\,$\AA. Remarkably, this estimate is applicable even in the limit where the
effective band dispersion is essentially that of free particles (e.g., for $\mu>4t$ in Fig. \ref{fig:whole-band}), despite the Sondheimer cancellation.
The results for the conductivities (inset of
Fig.~\ref{fig:whole-band}) can also be used to extract the components
of the Peltier tensor and to compute the Nernst coefficient via
Eq.~(\ref{eq:1Q_xy}).

Equation~(\ref{eq:1nu-estimate}) suggests that quasiparticles will
generally make a finite, typically negative contribution to the Nernst
coefficient. The magnitude of the effect is controlled by the product
$k_B T \tau / \hbar$. This can be small, e.g., when the scattering is
strong or at low temperatures in the impurity dominated regime, or
large, e.g.\ in clean systems with only moderate inelastic scattering.
In this latter regime the Boltzmann theory predicts that the range of
magnetic fields over which $Q_{xy}$ is linear diminishes, with the
crossover to $Q_{xy} \sim 1/B$ taking place at $\omega_c \tau \approx
1$. Although in practice this crossover between weak-field and
large-field regimes need not be sharp (or simple), in cases where such
behavior in $Q_{xy}(B)$ can be observed as the temperature is lowered,
it can serve as an independent evidence of coherent quasiparticle
transport.\cite{wdwu}

Up to now our discussion explicitly assumed that all relaxation
processes can be summarized by a single scattering time,
independent of the energy or momentum of the quasiparticles. This
simplified analysis can be readily generalized to three dimensions
[the factor of $1/2$ in Eq.~(\ref{eq:1mmW}) becomes $1/3$]. The
analysis in the next Section proceeds along these lines and
focuses on the contribution to the effect coming from the DDW
induced changes in Fermi surface, which is very much in the spirit
of the proposal of
Refs.~\onlinecite{DDW-proposal,sharp-signature,DDW-ARPES,DDW-infrared}.

We believe, however, that a more detailed modelling of relaxation may
be necessary, both in HTSC and other cases.  For example, we have
completely neglected issues such as the dependence of the scattering
time on energy or location on the Fermi surface, as well as the
difference in the relaxation of electrical and energy
currents~\cite{z2} (which may lead to different scattering times, as
may be the case for scattering by phonons). Clearly, these issues
cannot be addressed without a specific material in mind. We defer our
discussion of particular experiments until Sec.~\ref{sec:discussion}.
Here, we focus on the energy dependence of the scattering time, for
which a general statement is possible.

Consider, for simplicity, a nearly isotropic Fermi surface (this
corresponds to the regime $\mu > t$ in Fig.~\ref{fig:whole-band}). One
may then estimate
\begin{equation}
\Theta_H = \frac{a^2}{\ell_B^2} \, \frac{\tau(\mu)}{\hbar} \,
\frac{1}{2 \pi N(\mu) a^2},\label{eq:1thetaH}
\end{equation}
where $N(\mu)$ is the density of states at the Fermi surface. If the
scattering is primarily due to weak quenched disorder (or phonons above
the Debye temperature) the corresponding rate is approximated by $1 /
\tau (\mu) \propto N(\mu)$. Then, properly accounting
for the $\mu$ dependence of $\tau$ amounts to an additional factor of 2
in the estimate of the Nernst coefficient.~\cite{constell} More
generally, even for inelastic processes, one expects the scattering
rate to be an increasing function of the electronic density of states,
and therefore act to enhance the estimate in
Eq.~(\ref{eq:1nu-estimate}).

\section{The DDW state}
\label{sec:ddw}

The DDW state is specified by the following
Hamiltonian~\cite{DDW-proposal,sharp-signature,DDW-ARPES,DDW-infrared}
\begin{equation} \label{H-DDW}
H = \sum_{s} \int_{BZ} \frac{d^2 k}{(2 \pi)^2} \left(
\epsilon_\mathbf{k} c^\dagger_{\mathbf{k},s} c_{\mathbf{k},s} + i
\Delta_\mathbf{k} c^\dagger_{\mathbf{k},s} c_{\mathbf{k} + \mathbf{Q},
s} + \text{h.c.} \right) ,
\end{equation}
where $c^\dagger_{\mathbf{k},s}$ is the creation operator for a
quasiparticle with momentum $\mathbf{k}$ and spin $s$, and
$\epsilon_\mathbf{k}$ is the effective quasiparticle dispersion. When
present, the DDW order parameter gives rise to a potential,
$\Delta_{\mathbf{k}}$, connecting states separated by the ordering
wavevector $\mathbf{Q} = (\pi, \pi)$, creating two bands in a reduced
Brillouin zone. The eigenvalues of this Hamiltonian are
\begin{equation}
\epsilon_{\mathbf{k}}^\pm = \frac{1}{2} ( \epsilon_\mathbf{k} +
\epsilon_{\mathbf{k} + \mathbf{Q}} ) \pm \frac{1}{2} \sqrt{
(\epsilon_\mathbf{k} - \epsilon_{\mathbf{k} + \mathbf{Q}})^2 + 4
\Delta_\mathbf{k}^2 }.
\end{equation}
The essential ingredient in transport calculations is the particle
current operator. In the basis in which the Hamiltonian is diagonal,
the current operator is given by
\begin{equation} \label{eq:2current}
\mathbf{j} = - e \sum_s \int_{RBZ} \frac{d^2k}{(2 \pi)^2} \,
\chi^\dagger_{\mathbf{k}, s} \left( \begin{array}{cc}
\nabla_{\mathbf{k}} \epsilon^+_\mathbf{k} / \hbar &
\mathbf{v}^\mathrm{inter}_\mathbf{k} \\
\mathbf{v}^\mathrm{inter}_\mathbf{k} & \nabla_{\mathbf{k}}
\epsilon^-_\mathbf{k} / \hbar \end{array} \right) \chi_{\mathbf{k}, s} .
\end{equation}
The off-diagonal elements
\begin{equation}
\mathbf{v}^\mathrm{inter}_\mathbf{k} = \frac{1}{\hbar} \,
\frac{(\epsilon_\mathbf{k} - \epsilon_{\mathbf{k} + \mathbf{Q}})
\nabla_\mathbf{k} \Delta_\mathbf{k} - (\nabla_\mathbf{k}
\epsilon_\mathbf{k} - \nabla_\mathbf{k} \epsilon_{\mathbf{k} +
\mathbf{Q}}) \Delta_\mathbf{k}}{\sqrt{ (\epsilon_\mathbf{k} -
\epsilon_{\mathbf{k} + \mathbf{Q}})^2 + 4 \Delta_\mathbf{k}^2 }}
\end{equation}
result in interband contributions to transport. However, if the energy
gap to the second band is larger than $k_B T$ and $\hbar / \tau$, these
interband contributions may be neglected and the Boltzmann equation
recovered for dc transport (see Appendix for an example where interband
and intraband contributions must be treated on equal footing).

\begin{figure}
\includegraphics[width=3.25in]{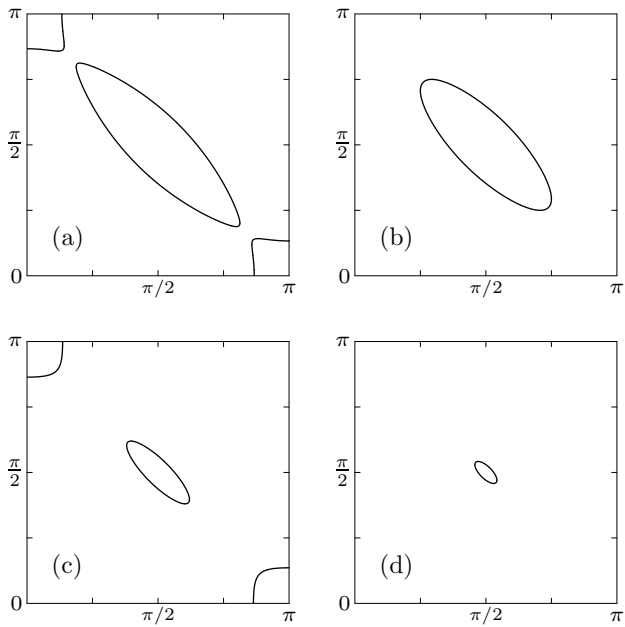}
\caption{\label{fig:FS} Fermi surface in the upper right quadrant of
the Brillouin zone, for parameter values representing the four regimes
described in the text: (a) ``weak'' regime ($\Delta = 0.05t$, $\mu =
-0.9t$), (b) ``moderate'' regime ($\Delta = 0.25t$, $\mu = -0.9t$), (c)
``ambipolar'' regime ($\Delta = 0.35t$, $\mu = -0.4t$), (d) ``Dirac''
regime ($\Delta = 0.75t$, $\mu = -0.2t$).}
\end{figure}

Following common practice
we approximate $\epsilon_\mathbf{k} = - 2 t ( \cos k_x + \cos k_y) + 4
t' \cos k_x \, \cos k_y$ and $\Delta_\mathbf{k} = \Delta (\cos k_x -
\cos k_y)$ (in this section we set $a = 1$). For the effective
band-structure parameters we shall use $t = 0.3\,$eV and $t' = 0.3 t$,
with the chemical potential in the range $-t \lesssim \mu \lesssim -
0.75 t$. In the absence of DDW this choice of parameters is consistent
with the character of the Fermi surface observed in ARPES~\cite{Rpez}
in the doping range of $5\% \sim 20\%$. The choice of $\Delta$ requires
some care as the effect of DDW order is sensitive to the filling.  For
clarity we loosely classify the different regimes as ``weak'',
``moderate'',``ambipolar'', and ``Dirac''.  The names are meant to
reflect qualitatively the character of the Fermi surface in each of the
regimes (see Fig.~\ref{fig:FS}).

The Dirac DDW occurs when $\Delta$ is large in the vicinity of
half filling. Its spectrum consists of Dirac points at $(\pm \pi /
2, \pi / 2)$. The ambipolar regime can occur at moderately strong
$\Delta$, in the vicinity of half filling. Here the Fermi surface
consists of three well formed sheets, two hole-like centered about
$(\pm\pi/2,\pi/2)$ and one electron-like centered about $(\pi,0)$,
and the system is thus composed of two types of carriers. We
discuss these two regime in the Appendix, as we do not believe
either of them is realized in the doping range where enhanced
Nernst effect is observed. The Fermi surfaces of both
of these states are sufficiently remarkable to be easily ruled out
based on the available ARPES data.~\cite{Rpez} Experimentally, 
neither well defined electron pockets in the antinodal direction
nor Dirac nodes (with the Fermi energy essentially at the node) are of 
relevance to the normal state of hole-doped cuprates.

In their stead, in this Section, we present results for DDW order
in the weak (high temperature) and moderate (low temperature)
regimes (for which we use, as representative values, $\Delta =
0.05t$ and $0.25t$, respectively). Qualitatively, the former
regime is where DDW order just begins to set in by disconnecting
the Fermi surface into two hole pockets [closed about $(\pm \pi/2,
\pi/2)$] and one electron pocket [closed about $(\pi,0)$]. Given
the rather short scattering times at these elevated temperatures
it is not clear whether such minute changes in the Fermi surface
geometry can be discerned (and motivated) from ARPES. However,
this regime is virtually unavoidable as one begins one's descent
into the pseudogap. It is worth remarking that since the
electron-like pocket is only such in its topology (and name), e.g.
its contribution to $\sigma_{xy}$ is hole-like, the effect of the
DDW order is expected to be rather insignificant. At larger values
of $\Delta$ and away from half filling, the regime with moderate
DDW order is intended as a caricature of the pseudogap at low
temperatures, near the superconducting $T_c$. Here the electron
pocket is absent and only two hole pockets
(``arcs''~\cite{DDW-ARPES}) remain. Perhaps this is the regime of
most interest, as the behavior in this regime provides much of the
motivation for introducing DDW order to explain the pseudogap in
the first place.

\begin{figure}
\includegraphics[width=3.25in]{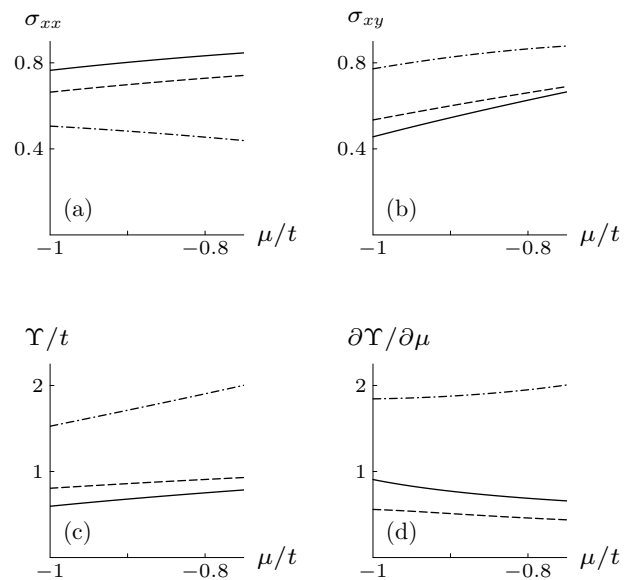}
\caption{\label{fig:transport} (a) Conductivity [in units of $(e^2 /
\hbar) (t \tau / \hbar)$], (b) Hall conductivity [in units of $(e^2 /
\hbar) (t \tau / \hbar)^2 (a^2 / \ell_B^2)$] (c) energy scale (or Hall
angle) $\Upsilon$, and (d) $\partial \Upsilon /
\partial \mu$ that enters the Nernst coefficient. All quantities are
given as a function of chemical potential (in the physically relevant
range) for $\Delta = 0$ (solid line), $\Delta = 0.05$ (dashed line),
and $\Delta = 0.25$ (dashed-dotted line). Note that a positive
$\partial \Upsilon / \partial \mu$ implies a negative Nernst
coefficient [see Eq.~(\ref{eq:1simplenu})].}
\end{figure}

Our results in these two regimes (as well as for $\Delta=0$) are
presented in Fig.~\ref{fig:transport}. Using Eqs.~(\ref{eq:1sigma-xx})
and~(\ref{eq:1sigma-xy}) we numerically calculated $\sigma_{xx}$ and
$\sigma_{xy}$, extracted the Hall angle (and $\Upsilon$), and evaluated
its derivative which appears in the result for the Nernst effect,
Eq.~(\ref{eq:1simplenu}). Even for the moderate regime, $\Delta = 0.25$,
the result for $\partial \Upsilon / \partial \mu$ [panel (d)] shows
that the overall enhancement in magnitude of the Nernst coefficient is
at most by a factor of 3. Moreover, the sign of the Nernst signal
(which is opposite to the sign of $\partial \Upsilon / \partial \mu$)
remains negative. These are the central results of this Section, and
are important for the comparison with experiment in
Sec.~\ref{sec:discussion}.

Two other points are perhaps worth noting in these results. First, as
anticipated, the effect of a weak DDW is very small (solid \emph{vs.}
dashed lines in Fig.~\ref{fig:transport}). Second, we note that for
moderate DDW $\sigma_{xx}$ is decreasing as $\mu$ is increasing, and
that this is the main source of the enhanced $\partial
\Upsilon/\partial \mu$. This is the beginning of the trend which
becomes especially pronounced in the Dirac regime of the model (see
Appendix).

\section{Experiments}
\label{sec:discussion}

\subsection{HTSC and DDW order}

We are now in a position to compare transport properties of the
DDW scenario with experimental data, obtained from different
underdoped HTSC.~\cite{Ong-nature,Ong-onset,Capan,Wen} 
Consider first the
high temperature regime, where the observed effect is small, with
a magnitude of about $25\,$nV/K$\,$T, usually negative, roughly
independent of temperature, and linear in the applied magnetic
fields over a very wide range. These experimental findings are
consistent with our discussion of Sec.~\ref{sec:1theory} (i.e.,
without DDW), where we argued for a generically negative sign of
the effect. To fit the actual magnitude of the effect with our
calculations requires taking $k_B T \tau / \hbar \approx 0.2$,
which is probably somewhat small for the cuprates. Additional
factors which may affect the Nernst signal and were not taken into
account may include the short scattering time of quasiparticles in
the antinodal region and a dependence of the scattering time on
energy (see below). A separate measurement of the contribution of
the two terms in Eq.~(\ref{eq:1Q_xy})~\cite{Ong-nature,Ong-onset}
reveals a significant Sondheimer cancellation in some cases (in
itself suggestive of the quasiparticle origin of the signal), but
not in others, indicating that these additional factors may be
more material dependent.

As the temperature is lowered, below an onset temperature which is well
separated from the superconducting transition temperature $T_c$, the
Nernst coefficient begins to increase. The signal becomes large and
positive, reaching a value of about $1\,\mu$V/K$\,$T near $T_c$
(depending on material, the increase in magnitude is typically by a
factor of 50--100). There is no Sondheimer cancellation in this case,
as all of the increase arises from the first term in
Eq.~(\ref{eq:1Q_xy}), while the second goes down to zero. Deviations
from linearity in the magnetic field dependence are observed at
moderate fields in some samples. Below $T_c$ the line-shapes are no
longer linear at small fields as one enters the flux flow
regime.~\cite{Flux-flow}

The question is then whether the enhancement of the Nernst signal above
$T_c$ can be ascribed to DDW order. The results of Sec.~\ref{sec:ddw}
make such prospect unlikely. First, the Nernst coefficient we find when
the DDW order is introduced remains \emph{negative}, opposite in sign
to the effect in experiment. Second, the enhancement in its magnitude
due to modifications of the Fermi surface by the DDW potential is
rather modest, a factor of three at most. The Nernst effect in the DDW
phase is therefore not fundamentally different than what is expected
from general notions of quasiparticle transport. Given that $T \tau$
does not change appreciably in the relevant temperature regime (see
below) the overall magnitude of the predicted signal is significantly
smaller than $1\,\mu$V/K$\,$T.

At this point it is important to consider whether this conclusion may
be modified by changing any of the underlying assumptions. We have
investigated to some extent the dependence on different choices of
effective band parameters (beyond those reported in this paper), with
no change to our conclusions. In addition to band structure, it is
important to consider the role of scattering. Indeed, the magnitude of
the Nernst signal depends on $T \tau$. This quantity may increase by a
factor of 3 in the pseudogap regime as inferred from Hall
data~\cite{OngHall} (it is independent of temperature according to the
conductivity data\cite{qpdamping}). We note that even if the scattering
time increases dramatically when the temperature is lowered, the
magnitude of the Nernst signal will increase, but the sign will not
change.

Another assumption  we have made
is to ignore the contribution arising from the scattering time's dependence
on energy. This contribution may be of either sign, and, in particular,
if $\partial \tau / \partial \mu < 0$ there is an additional
contribution to the Nernst signal which is of positive sign. Can this
contribution be so strong so as to overwhelm the band structure
contribution and lead to a large positive signal? While unlikely, we
note that this assumption leads to other discrepancies with the data.
In particular, this would require a large contribution to the second
term in Eq.~(\ref{eq:1Q_xy}) (equal to one half of the contribution to
the first term). However, clearly this is not seen in the
experiment.~\cite{Ong-nature,Ong-onset}

\subsection{Other Nernst experiments}

Our simple estimate in Sec.~\ref{sec:1theory},
Eq.~(\ref{eq:1nu-estimate}), suggests that the Nernst effect due to
quasiparticles may be quite large if $k_B T \tau/\hbar \gg 1$. Why is
it, then, that large Nernst coefficients are not typically observed
in conventional metals,
where $k_B T \tau/\hbar$ is expected to grow as
the system becomes more coherent, its magnitude limited only by the
sample's purity?
Much of the data in metals\cite{metal-data} is
collected in the temperature regime dominated by classical phonons. Here, the resistivity varies linearly with temperature, while the Nernst signal does not show a strong dependence on temperature
(with a few notable exceptions, e.g., Ni).
There is a considerable variation in the magnitude of the Nernst signal (e.g., $\nu = - 0.5\,$nV/K$\,$T in Pb, $\nu = 120\,$nV/K$\,$T in Nb).
In this regime oftentimes $k_B T \tau/\hbar < 1$,~\cite{allen} making our simple estimate in Sec.~\ref{sec:1theory}
broadly consistent with the data (although not a substitute for a detailed material-specific modeling).
The situation at lower temperatures is less
clear, even if we restrict our attention to electron-phonon scattering
(so that the resistivity varies as $T^5$). In this regime electric and
heat currents have different relaxation rates (as deduced from
electrical and thermal conductivities), leading to a violation of the
Wiedemann-Franz law.~\cite{z2} This difference
might lead to a large conductivity in the denominator of
Eq.~(\ref{eq:1Q_xy}), and hence a suppression of the Nernst signal.

Following the discovery in the hole-doped cuprates, relatively large
Nernst signals have been documented in other
strongly correlated materials.~\cite{ambipolar,kondo,Balci,weida}
Perhaps the most obvious difference between these and conventional materials, as far as the Nernst effect is concerned,
is in the strong temperature dependence of the observed signal.
Since in at least some of these materials the transport is apparently
due to quasiparticles, it is of interest to consider the corresponding
Nernst measurements
in the light of our results.
While far from a complete theory
the discussion of Sec.~\ref{sec:1theory} does allow for
a few general observations.

In NbSe$_2$ (for $7\,$K $\lesssim T \lesssim 60\,$K),~\cite{ambipolar}
the Nernst signal is relatively large and negative
(except very close to $T_c$),
and hence is naturally attributed to quasiparticles.
The Nernst
coefficient reaches a maximum, $\nu \approx -0.12 \, \mu$V/K$\,$T, at $T
\approx 20\,$K, roughly the same temperature at which the Hall number passes
through zero. The behavior of the Hall number is apparently sensitive to the charge density wave induced reconstruction of the Fermi surface,
which occurs below $T_{\rm{CDW}}=32.5$K.
As recognized by the authors of Ref.~\onlinecite{ambipolar}, $R_H=0$
suggests an ambipolar origin of the signal due to bands of oppositely
charged carriers, reminiscent of that encountered in
semiconductors.~\cite{Delves} We note, however, that a simple ambipolar
picture leads to a positive Nernst signal (see Appendix), opposite to
the experimentally observed signal.
It is also interesting to note that while the Hall number remains essentially constant for $T>T_{CDW}$
the Nernst coefficient grows appreciably (up to 75 percent of its max
value). This suggests to us that the enhanced Nernst
coefficient here owes its existence as much (if not more) to enhanced
coherence in the system as to an ambipolar compensation.
Alternately, perhaps, the Nernst measurement is sensitive to incipient CDW order, more so than $R_H$.\cite{AJM}

Yet a larger Nernst signal was reported very recently in the normal state
of CeCoIn$_5$, a heavy fermion compound.\cite{kondo} Here, the Nernst signal is
also negative
and can be as high as $2\,\mu$V/K$\,$T at low temperatures, suggestive
of a highly coherent state (large $\tau$). As discussed in Sec.~\ref{sec:1theory}, an
increased coherence is expected to be accompanied by a reduction of the
linear response regime (in magnetic field). Indeed, the measurements in
CeCoIn$_5$ are highly suggestive of such behavior.~\cite{foot:CECOIN}

\section{Conclusions}

The pseudogap is typically envisaged as a strongly fluctuating
crossover (possibly quantum critical) regime where
superconductivity and possibly other orders are present in some
form, yet not fully condensed. Theories implementing this
intuitive picture are necessarily subtle and complex (and rarely
applicable directly to experiments). The DDW proposal is a breath
of fresh air in that regard, here the pseudogap phase is a true
phase of matter where the standard crisp notions of order
parameter and quasiparticles apply and can be used to make
predictive statements. In particular, most of the effects of DDW
order discussed to date can be linked directly to the changes of
the Fermi surface geometry due to the breaking of the
translational symmetry. Implications of such Fermi surface
reconstruction for thermodynamic and dynamic properties of the
system (including transport) were addressed in the past (see, e.g.
Ref~\onlinecite{sharp-signature}) and found consistent with the
observed phenomenology.

In this work we have extended this investigation, focusing in
particular on the Nernst effect. Our main result is that
quasiparticle transport inside the DDW phase cannot explain the
Nernst phenomena observed in hole-doped cuprates.

This conclusion is based on a Boltzmann theory based
calculation.~\cite{collective} The Nernst coefficient has a
contribution originating from the changes of the Fermi surface shape,
which we compute explicitly, and one from the changes in the scattering
rate (treated as a phenomenological input). The first contribution
alone predicts a negative Nernst coefficient, which is in contradiction
with the experiment. It is possible that upon inclusion of the second
term the overall sign can be reversed, however, we presented arguments
for why the overall magnitude cannot approach the experimentally
observed signal. 
We believe, therefore, that the physics of large
Nernst effect in the cuprates lies elsewhere.

Our general analysis can be applied, albeit qualitatively, to the
recent experiments of Bel \emph{et al.} on NbSe$_2$ and CeCoIn$_5$. It
suggests that the data in these materials is not inconsistent with a
quasiparticle based interpretation, although a more detailed analysis
is required. Likewise, it would be interesting to examine other cases
(including, e.g., conventional density waves and other strongly
correlated systems) along similar lines.

\begin{acknowledgements}
We thank Kamran Behnia, Paul Chaikin, Sudip Chakravarty, Eduardo Fradkin, 
Leonid Glazman, David Huse, Steve Kivelson, Andy Millis, Phuan Ong, Weida Wu,
and especially Shivaji Sondhi for numerous enlightening discussions
and for their valuable input. This research was supported by the NSF
through grants EIA 02-10736 (IU), DMR 99-78074, DMR 02-13706  
and by the David and Lucile Packard foundation.
\end{acknowledgements}

\appendix*

\section{DDW order and particle-hole symmetry}
\label{sec:ambinodal}

The discussion of DDW order in Sec.~\ref{sec:ddw} was limited by its
potential relevance to the cuprates. In this section, we explore other
regimes of the DDW Hamiltonian, Eq.~(\ref{H-DDW}), that may be of
broader interest (they may also arise in conventional density waves).
Particle-hole symmetry (either exact,  $\sigma_{xy} = \alpha_{xx} = 0$,
or approximate) plays an important role in both of the regimes
considered below.

\subsection{Ambipolar DDW}
\label{sec:ambipolar}

At moderately strong DDW order and in the vicinity of half filling (we
take here $\Delta = 0.35t$ and $\mu = -0.4t$ as representative values),
the DDW model produces a well defined electronic pocket at $(\pi, 0)$,
in addition to the hole pockets at $(\pm \pi/2, \pi/2)$. [see Fig.
\ref{fig:FS}(c)]. The Nernst coefficient in this regime is positive and
modestly enhanced. For example, for the parameters above, $\partial
\Upsilon / \partial \mu = -1.99$. The reason for this behavior may be
traced to the existence of two types of carriers in the system.

As a representative model of this behavior consider two oppositely
charged but otherwise identical species of free carriers. We then have
\begin{equation}
\sigma_{xx} = (n_h + n_e) \frac{e^2 \tau}{m}, \qquad \sigma_{xy} = (n_h
- n_e) \frac{e^2 \tau}{m} \omega_c \tau,
\end{equation}
leading to a Nernst coefficient
\begin{equation} \label{eq:ambipolar}
\nu = \frac{2 \pi^2}{3} \frac{k_B^2 T}{e B}  \frac{n_e n_h}{(n_e +
n_h)^2} \left( \frac{1}{\epsilon^e_F} + \frac{1}{\epsilon_F^h} \right)
\omega_c \tau ,
\end{equation}
where $\partial n_e / \partial \mu = n_e / \epsilon_F^e$, appropriate
for two dimensions, was used (and similarly for the hole band). This
ambipolar Nernst signal
is maximal when the bands are exactly compensated, and is positive.
Substituting $n_e = n_h = k_F^2 / 4 \pi$ and rewriting
\begin{equation} \label{kFellB}
B\nu=\frac{2 \pi^2}{3} \, \frac{k_B}{e} \, \frac{k_B T \tau}{\hbar} \,
\frac{1}{ (k_F \ell_B)^2},
\end{equation}
we arrive at an expression which can be compared against the lattice
result in Eq.~(\ref{eq:1simplenu}). Provided we loosely identify
$k_F\sim 1/a$, the comparison suggests that the ambipolar Nernst effect
need not be particularly larger then a signal from a single band of
carriers: long scattering times are essential for either
Eq.~(\ref{eq:1simplenu}) or Eq.~(\ref{kFellB}) to produce a substantial
Nernst coefficient. In situations when $k_F$ and $a$ are
unrelated Eq.~(\ref{kFellB}) may lead to significant enhancements of
Nernst signals (as may be the case of Fermi pockets associated with
spin-density waves~\cite{AJM}).

\subsection{Dirac DDW}
\label{sec:nodal-I}

Near half-filling and with strong DDW order ($\Delta > 0.6t)$,
the dispersion around the Fermi surface approaches that of a Dirac
particle, with dispersion $\epsilon_\mathbf{k}^\pm = \pm \hbar
\sqrt{(v_F k_x)^2 + (v_\Delta k_y)^2}$, where the momenta here are
measured relative to $(\pm \pi/2,\pi/2)$ along the diagonals of the
Brillouin zone. The density of states of the Dirac Hamiltonian in two
dimensions vanishes at the node, $N (\epsilon) = |\epsilon|/(2 \pi
\hbar^2 v_x v_y)$, giving rise to a host of unusual properties. To
simplify the discussion we shall consider an isotropic case, $v_x = v_y
= v$.

Provided the chemical potential (or scattering time) is sufficiently
large ($\mu \tau > \hbar$) we can still neglect interband scattering,
and use Boltzmann theory in the weak field regime. We then have
\begin{equation}
\sigma_{xx} = \frac{e^2 \tau}{4 \pi \hbar^2} \, |\mu|, \qquad
\sigma_{xy} = \frac{e^3 B \tau^2 v^2}{4 \pi \hbar^2 c} \mathrm{sign}(-
\mu).
\end{equation}
Here the dissipative conductivity is proportional to the density of
states at the Fermi energy (hence the factor of $|\mu|$), while the
Hall conductivity can be obtained by multiplying the conductivity by
the Hall angle
\begin{equation} \label{DiracHall}
\Theta_H = \omega_c \tau \mathrm{sign} (-\mu) = \frac{e B \tau}{m^{*}
c} \mathrm{sign} (- \mu) = - \frac{e B v^2 \tau}{\mu c},
\end{equation}
where $m^{*}=\hbar k_F/v$. As a result $\sigma_{xy}$ is constant on
either side of the node. Finally, the Nernst coefficient
\begin{equation}
\nu=- \frac{\pi^2}{3} \, \frac{k_B^2 T v^2}{c} \,
\frac{1}{\mu^2}
\end{equation}
mirrors the divergence of the Hall angle (as $\mu\rightarrow 0$) [cf.\
Eq.~(\ref{eq:1nernst1})]. As discussed in Sec.~\ref{sec:1theory}, this
enhancement is necessarily accompanied by a reduced regime of linear
response in magnetic field (with $Q_{xy}$ decreasing with magnetic
field for $\omega_c \tau \gtrsim 1$). It is also interesting to note
that for the linearized Dirac spectrum used here $\alpha_{xy}=0$ and
the large Nernst coefficient comes entirely from the enhancements in
the second term in Eq.~(\ref{eq:1Q_xy}) due to reduced conductivity.

Eventually, for $\mu \tau< \hbar$, the Boltzmann equation is no longer
applicable, and interband contributions to transport become important
[see Eq.~(\ref{eq:2current})]. Instead, a careful evaluation of Kubo
formulae treating the interband and intraband terms on equal footing is
necessary.~\cite{nodal-DDW,IUVOnodal}
Qualitatively, the singularities of the conductivity tensor at $\mu
= 0$ implied by the Boltzmann analysis above become smoothed. In
particular, the Hall angle [Eq.~(\ref{DiracHall})] rapidly changes its
value as a function of $\mu$, passing through 0 at $\mu = 0$,
consistent with the particle-hole symmetry at the node. The Nernst
coefficient at the node may then become large and positive [cf.\
Eq.~(\ref{eq:1nernst1})], arising entirely
from an unusually large\cite{IUVOnodal} $\alpha_{xy}$ and
``universally''\cite{fisherfradkin,nodal-DDW} small $\sigma\sim
e^2/\hbar$,
\begin{equation} \label{nodal-II-nernst}
B \nu = c_1 \frac{k_B}{e} \, \frac{k_B T \tau}{\hbar} \, \frac{B e v^2
\tau^2}{c \hbar} = c_1 \frac{k_B}{e} \, \frac{k_B T \tau}{\hbar} \left(
\frac{\epsilon_0  \tau}{\hbar} \right)^2,
\end{equation}
where $c_1$ is a numerical constant. The energy $\epsilon_0$ can be
identified as the energy gap between two lowest Landau levels of the
Dirac Hamiltonian (in the absence of disorder).

Equation~(\ref{nodal-II-nernst}) seemingly opens the door to a
positive Nernst signal which can become very large in
a clean system (as $\nu \propto \tau^3$).~\cite{Maki}
Here, we
note that the possibility of observing a large Nernst signal is very
restricted (even if $\tau$ is large). First, the chemical potential
must happen to be very close to the node, $|\mu| \lesssim \hbar /
\tau$. Second, assuming $\mu = 0$, a Nernst signal which is linear in
magnetic field will only be observed provided $\epsilon_0 \lesssim
\hbar / \tau$. In other words, in a very clean system, the Nernst
signal as a function of magnetic field will have a large initial slope,
and will very rapidly reach its maximal value of $Q_{xy} \approx k_B^2
T \tau / e \hbar$ (as in the case of Boltzmann results above).


\begin{thebibliography}{}
\bibitem{dmu}
The derivative with respect to $\mu$ arises from the low temperature
expansion of an integral over energy of the form $\int d \epsilon \, (-
\partial f / \partial \epsilon) \, \epsilon \ldots$ (where $f$ is the
Fermi-Dirac distribution function). It is important to bear in mind the
origin of this derivative when considering possible sources of $\mu$
dependence. For example, should the effective band structure parameters
have an explicit $\mu$ dependence, it is to be ignored for the purposes of
computing $\alpha$.
\bibitem{vanHove}
There is one notable exception to this estimate. At a special van
Hove filling the Fermi surface changes its topology, from electron-like to
hole-like. For the parameters used in Fig.~\ref{fig:whole-band} this
happens at $\mu = -1.2 t$. The value of $\partial \Upsilon / \partial
\mu$ diverges at this filling, and deviates significantly from 1 in a
very narrow range in its vicinity. We caution, however, that the
Boltzmann theory
fails at the van Hove filling due to the vanishing of the Fermi velocity at the saddle points of the bandstructure.
Also, in two dimensions, the single particle density of states is divergent at the van Hove filling which likely amplifies interaction effects.
Altogether, this leaves the question of the
magnitude of $\nu$ near the van Hove filling open.
\bibitem{constell}
A somewhat different situation can arise if the mean free path of
quasiparticles is fixed and independent of their momentum, e.g., due to
scattering off strong impurities.~\cite{AJM} In this case $1 / \tau
(\mathbf{k}) = v_\mathbf{k}/\ell$, where $\ell$ is the average distance between impurities.
This can lead to an
enhanced Nernst signal in two dimensions.
\bibitem{Flux-flow}
Below $T_c$, for weak magnetic field, the vortices are pinned and $\nu=0$.
When the magnetic field is
increased above a threshold, the Nernst signal $Q_{xy}$ develops and quickly
becomes large and positive as the vortices become more dense and easier to depin.\cite{Ong-nature}
A similar non-linearity is expected for
low fields in the dependence of the measured electric field on
temperature gradient; above a certain threshold, the temperature
gradient will be sufficiently strong to depin the
vortices.~\cite{Huebener}
\bibitem{foot:CECOIN}
The case of CeCoIn$_5$ is subtle, partly because the application of
magnetic field likely affects not only conduction electrons but also
(perhaps more strongly) their scatterers, e.g., localized moments, as
suggested by the magnetic field dependence of thermopower below
$10\,$K. At temperatures above $10\,$K the situation is clearer, to the
extent that the Nernst coefficient is still large, $\sim
0.5\,\mu\,$V/K$\,$T, the Hall thermopower is non-linear, while the
thermopower changes only by about $20\% $ in the entire range of
magnetic fields up to $12\,$T.
\bibitem{collective}
We have not explicitly considered the effects of collective behavior of the DDW order, 
which is not expected to modify our conclusions. The DDW order is not charged, and provided its collective modes
remain gapped they will presumably only renormalize the effective parameters of
the model and appropriate scattering times. The latter were treated by
us as phenomenological inputs.
\end{thebibliography}
\end{document}